\documentclass[prl,twocolumn,showpacs,aps,floatfix,superscriptaddress]{revtex4}
\usepackage{amsmath,amssymb,eucal,graphicx,bm}
\begin{document}
\title{Slow Kinetics of Brownian Maxima}
\author{E.~Ben-Naim}
%\email{ebn@lanl.gov}
\affiliation{Theoretical Division and Center for Nonlinear Studies,
Los Alamos National Laboratory, Los Alamos, New Mexico 87545}
\author{P.~L.~Krapivsky}
\affiliation{Department of Physics, Boston University, Boston,
Massachusetts 02215}
%\email{paulk@bu.edu}
\begin{abstract}
  We study extreme-value statistics of Brownian trajectories in one
  dimension. We define the maximum as the largest position to date and
  compare maxima of two particles undergoing independent Brownian
  motion. We focus on the probability $P(t)$ that the two maxima
  remain ordered up to time $t$, and find the algebraic decay $P\sim
  t^{-\beta}$ with exponent $\beta=1/4$. When the two particles have
  diffusion constants $D_1$ and $D_2$, the exponent depends on the
  mobilities, $\beta=\frac{1}{\pi}\arctan\sqrt{D_2/D_1}$.  We also use
  numerical simulations to investigate maxima of multiple particles in
  one dimension and the largest extension of particles in higher
  dimensions.
\end{abstract}
\pacs{05.40.Jc, 05.40.Fb, 02.50.Cw, 02.50.Ey}
%05.40.Jc       Brownian motion
%05.40.Fb       Random walks and Levy flights
%02.50.Cw       Probability theory
%02.50.Ey       Stochastic processe
\maketitle

\begin{figure}[t]
\includegraphics[width=0.4\textwidth]{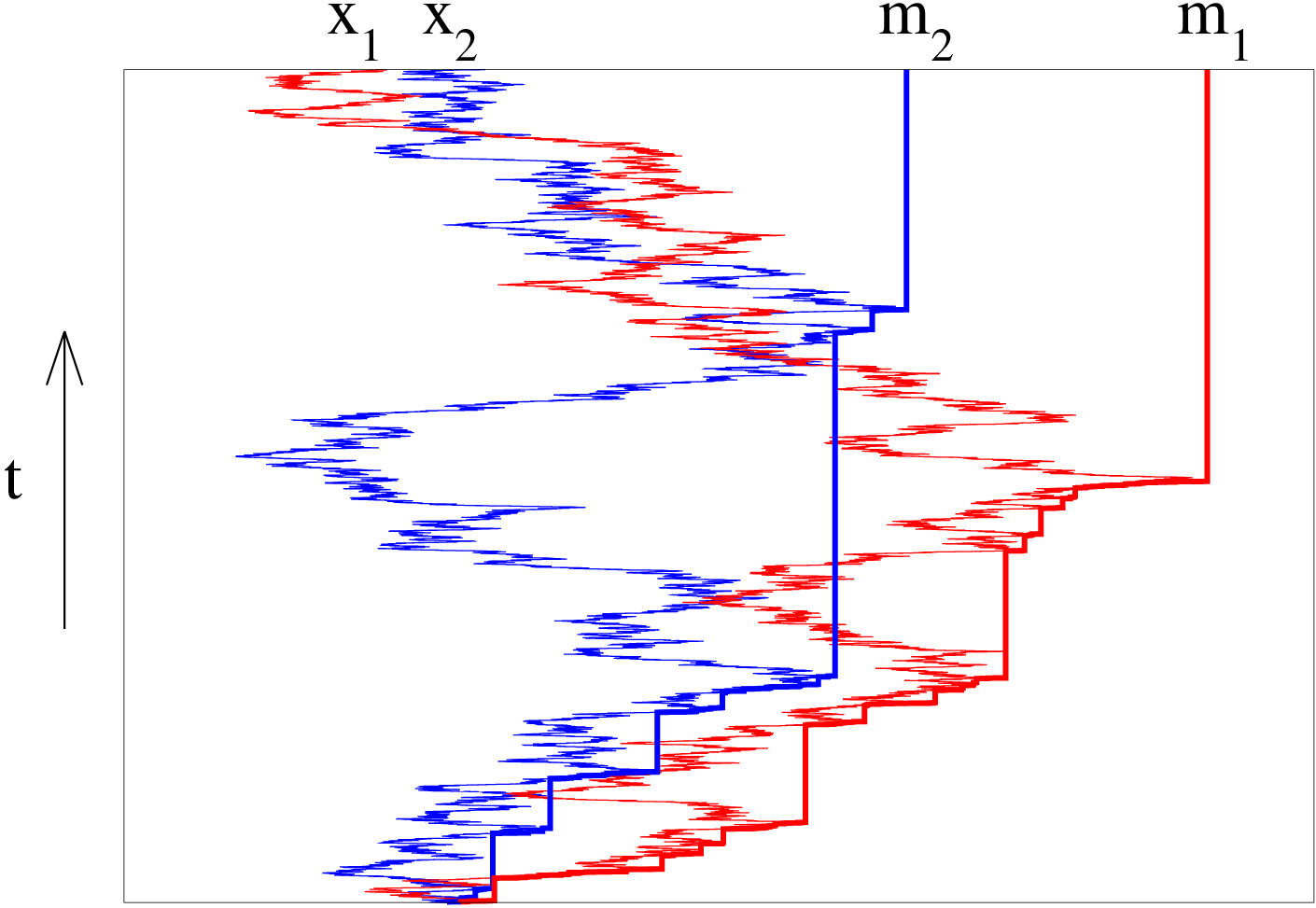}
\caption{Space-time diagram of the positions (thin lines) and the
ordered maxima (thick lines) of two Brownian particles.}
\label{fig-xt}
\end{figure}

Consider a pair of particles undergoing independent Brownian motion in
one dimension \cite{mp}.  These two particles do not meet with
probability that decays as $t^{-1/2}$ in the long-time limit. This
classical first-passage behavior holds for Brownian particles with
arbitrary diffusion constants. It holds even for particles undergoing
symmetric L\'evy flights \cite{sa,wf}, and has numerous applications
\cite{wf,sr}. Here, we generalize this ubiquitous first-passage
behavior to {\em maxima} of Brownian particles. Figure \ref{fig-xt}
shows that the maximal position of each particle forms a staircase and
it illustrates that unlike the position, the maximum is a
non-Markovian random variable \cite{pl,im}. We find that two such
staircases do not intersect with probability $P$ that is inversely
proportional to the one-fourth power of time, $P\sim t^{-1/4}$, in the
long-time limit. If the particles move with diffusion constants $D_1$
and $D_2$, the two maxima remain ordered during the time interval
$(0,t)$ with the slowly-decaying probability
\begin{equation}
\label{beta-gen}
P\sim t^{-\beta}\qquad\text{where}\qquad
\beta=\frac{1}{\pi}\arctan\sqrt{\frac{D_2}{D_1}}\,.
\end{equation}
In this letter, we obtain this result analytically and investigate
numerically related problems involving multiple maxima and diffusion
in higher dimensions.

Anomalous relaxation with nontrivial persistence exponents
\cite{msb,dhz,to}, enhanced transport due to disorder \cite{csg,bk},
and anomalous diffusion due to exclusion \cite{lkb,bs} are dynamical
phenomena that were recently demonstrated in experiments involving
Brownian particles. Understanding the nonequilibrium statistical
physics of these diffusion processes is closely intertwined with the
characteristic behavior of extreme fluctuations and the statistics of
extreme values \cite{krb,bms,dls,smcrf,mz,bk14}.

We first establish Eq.~\eqref{beta-gen} for two Brownian particles
having the same diffusion constant $D$. Let us denote the positions of
the particles at time $t$ by $x_1(t)$ and $x_2(t)$, and without loss
of generality, we assume $x_1(0)>x_2(0)$.  We define the maximum of
the first particle, $m_1(t)$, to be its rightmost position up to time
$t$; similarly, $m_2(t)$ is the maximal position of the second
particle.  Our goal is to find the probability $P(t)$ that the two
maxima remain ordered $m_1(\tau)> m_2(\tau)$ for all $0\leq \tau\leq
t$.

\begin{figure}[t]
\includegraphics[width=0.4\textwidth]{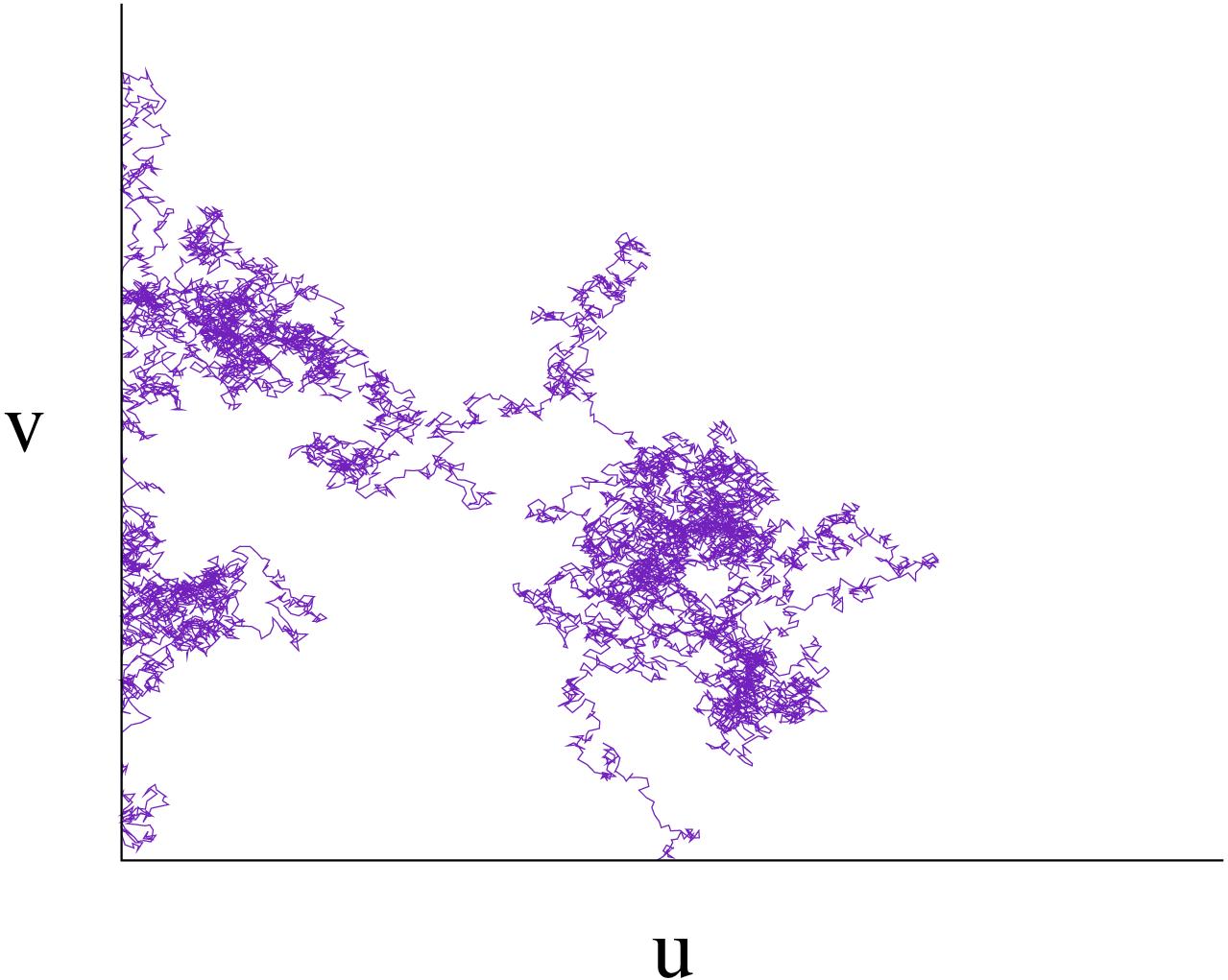}
\caption{Sample trajectory of the composite particle. There is upward
drift along the boundary $u=0$ and the boundary $v=0$ is absorbing.}
\label{fig-quad}
\end{figure}

The two maxima remain ordered {\em if and only if} \hbox{$m_1(\tau)>
x_2(\tau)$} at all times $0\leq \tau\leq t$.  Hence, to find $P$,
there is no need to keep track of the maximum $m_2$, and it suffices
to consider only the position $x_2$.  As a further simplification, we
focus on the {\em distance} of each particle from the maximum $m_1$
and introduce the variables
\begin{equation}
\label{uv-def}
u=m_1-x_1\qquad {\rm and}\qquad v=m_1-x_2.
\end{equation}
By definition, both distances are positive, $u\geq 0$ and \hbox{$v\geq
0$}.  The transformation \eqref{uv-def} maps the four variables (two
positions and two maxima) onto the two relevant variables (two
distances). Since the positions $x_1$ and $x_2$ undergo simple
diffusion, the distances $u$ and $v$ also undergo simple diffusion in
the domain \hbox{$u>0$} and \hbox{$v>0$}. Hence, the probability
density $\rho(u,v,t)$ obeys the diffusion equation \hbox{$\partial_t
\rho = D\nabla^2 \rho $} with \hbox{$\nabla^2=\partial^2_{u}
+\partial^2_{v}$} along with the boundary conditions
$\rho\big|_{v=0}=0$ and \hbox{$(\partial_u -
\partial_v)\rho\big|_{u=0}=0$}.  The boundary $v=0$ is absorbing so
that position $x_2$ does not exceed maximum $m_1$.  The second
boundary condition (see Supplemental Material for derivation) is more
subtle, and it effectively implies upward drift along the boundary
$u=0$: When the maximum increases, $m_1\to m_1+\delta m$, one distance
remains the same, $u=0$, but the second distance increases, $v\to
v+\delta m$ (Fig.~\ref{fig-quad}).

The probability $P(t)$ is the integral of the probability density,
$P(t)=\int_0^\infty \int_0^\infty du\,dv\, \rho(u,v,t)$.  This
quantity equals the survival probability of a ``composite'' particle
with coordinates $(u,v)$ that is undergoing Brownian motion in two
dimensions. This composite particle starts somewhere along the
boundary $u=0$, and it diffuses in the domain $u>0$ and $v>0$. The
particle experiences drift along the boundary $u=0$ but it is
annihilated when it reaches the boundary $v=0$ (Figure
\ref{fig-quad}).

In general, the probability $P$ depends on the initial coordinates
$u(0)$ and $v(0)$. It is convenient to compute the probability $P$
directly rather than through the probability density $\rho(u,v,t)$.
With the shorthand notations $X\equiv u(0)$ and $Y\equiv v(0)$, the
probability $P\equiv P(X,Y,t)$ obeys the standard diffusion equation
\cite{ghw,bk-mult}
\begin{equation}
\label{S-eq}
\frac{\partial P}{\partial t} = D\nabla^2 P 
\end{equation}
with the Laplace operator
\hbox{$\nabla^2=\partial^2_{X}+\partial^2_{Y}$}.  The initial
condition is $P=1$ in the region \hbox{$X\geq 0$} and \hbox{$Y>0$},
and the boundary conditions are \hbox{$P\big|_{Y=0}=0$} and
\hbox{$(\partial_{X}+\partial_{Y})P\big|_{X=0}=0$}. The former
reflects that the boundary $Y=0$ is absorbing, and the second is a
consequence of the drift (see Supplemental Material for details). Our
problem corresponds to the special case $X=0$ and
\hbox{$Y=x_1(0)-x_2(0)$}.

In terms of the polar coordinates \hbox{$R=\sqrt{X^2+Y^2}$} and
\hbox{$\theta=\arctan(Y/X)$}, the probability $P\equiv P(R,\theta,t)$
obeys the diffusion equation \eqref{S-eq} with the Laplace operator
\begin{equation*}
\nabla^2=
\frac{\partial^2}{\partial R^2} 
+\frac{1}{R} \frac{\partial}{\partial R }
+\frac{1}{R^2}\frac{\partial^2}{\partial \theta^2}
\end{equation*}
The first boundary condition is simply
$P\big|_{\theta=0}=0$. The second boundary condition
becomes
\begin{equation}
\label{bc-polar}
\left(R\,\frac{\partial P}{\partial R}-\frac{\partial P}
{\partial \theta}\right)
\,\Big|_{\theta=\pi/2}=0
\end{equation}
where we have utilized
$\partial_X=\cos\theta\,\partial_R-R^{-1}\sin\theta\,\partial_\theta$
and
\hbox{$\partial_Y=\sin\theta\,\partial_R+R^{-1}\cos\theta\,\partial_\theta$}.

\begin{figure}[t]
\includegraphics[width=0.45\textwidth]{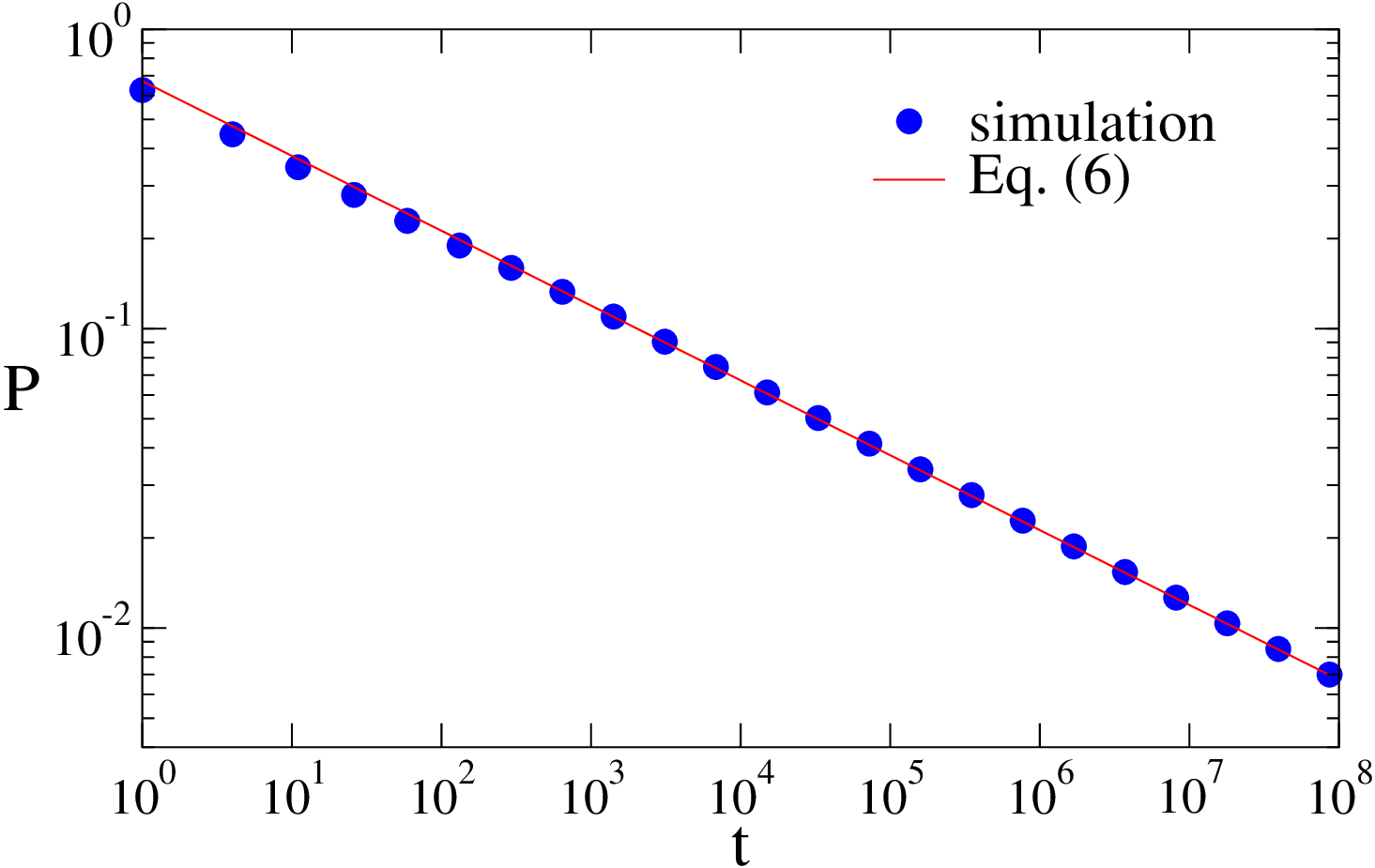}
\caption{The probability $P$ versus time $t$. Shown are Monte Carlo
  simulation results, obtained from $10^8$ independent realizations
  (circles).  Also shown as reference (line) is the theoretical
  prediction \eqref{Pt}.}
\label{fig-st}
\end{figure}

In the long-time limit, the solution to \eqref{S-eq} has a separable
form \cite{bk-mult}
\begin{equation}
\label{P-sol}
P(R,\theta,t)\sim \left(\frac{R^2}{Dt}\right)^{\beta} 
\sin \big(2\beta\theta\big).
\end{equation}
This form can be conveniently obtained using dimensional analysis: the
probability $P$ is dimensionless and the quantity $R^2/(Dt)$ is the
only dimensionless combination of the variables $R, D, t$. Hence, we
anticipate \hbox{$P(R,\theta,t)\sim (R^2/Dt)^\beta
f(\theta)$}. Plugging this expression into \eqref{S-eq} we see that
the left-hand side vanishes in the long-time limit, and the function
$f$ obeys $f''+(2\beta)^2f=0$.  We choose
$f(\theta)=\sin\left(2\beta\theta\right)$ to satisfy the boundary
condition $S\big|_{\theta=0}=0$.  Next, we substitute \eqref{P-sol}
into \eqref{bc-polar}, and observe that the second boundary condition
is obeyed when $\tan (\beta\pi)=1$.  Thus $\beta=1/4$ and we arrive at
the slow kinetics (Fig.~\ref{fig-st})
\begin{equation}
\label{Pt}
P\sim t^{-1/4}.
\end{equation}
Importantly, the decay exponent is an eigenvalue of the {\em angular}
component of the Laplace operator, and it is specified by the boundary
conditions. We note that the behavior \eqref{Pt} also characterizes
the probability that a particle diffusing on a plane avoids a
semi-infinite needle \cite{cr}.

Consider now the general case where the two particles have diffusion
constants $D_1$ and $D_2$.  The transformation $(x_1,x_2)\to
(\widehat{x}_1, \widehat{x}_2)$ with \hbox{$(\widehat{x}_1,
\widehat{x}_2) = (x_1/\sqrt{D_1},x_2/\sqrt{D_2})$} maps this
anisotropic Brownian motion onto isotropic Brownian motion in two
dimensions.  The maxima are also rescaled, \hbox{$(m_1,m_2)\to
(\widehat{m}_1,\widehat{m}_2)$} with
\hbox{$(\widehat{m}_1,\widehat{m}_2)=(m_1/\sqrt{D_1},m_2/\sqrt{D_2})$}. The
two maxima remain ordered, $m_1>m_2$, as long as
\begin{equation}
\label{condition}
\sigma\,\widehat{m}_1 > \widehat{m}_2\qquad \text{with}
\qquad\sigma=\sqrt{D_1/D_2}.
\end{equation}
Thus, we expect that the exponent depends on the ratio of diffusion
constants, $\beta\equiv \beta(D_1/D_2)$. When one particle is
immobile, the problem simplifies.  If $D_1=0$, the maxima remain
ordered if the particles do not meet, $x_1(0)>x_2(t)$, and hence
$P\sim t^{-1/2}$.  In the complementary case $D_2=0$, an immobile
particle can not overtake a maximum set by an mobile particle and
$P=1$.  The limiting values are therefore $\beta(0)=1/2$ and
$\beta(\infty)=0$, and since the exponent should be a monotonic
function of the ratio $D_1/D_2$, we deduce $0\leq \beta\leq 1/2$.

The above analysis is straightforward to generalize if instead of
\eqref{uv-def} we use the distances
\hbox{$u=\widehat{m}_1-\widehat{x}_1$} and
\hbox{$v=\sigma\,\widehat{m}_1-\widehat{x}_2$}.  Again, diffusion
takes place in the domain $u>0$ and $v>0$, and the boundary conditions
are $P|_{Y=0}=0$ and
\hbox{$(\partial_{X}+\sigma\,\partial_{Y})P\big|_{X=0}=0$}.  In polar
coordinates, the latter boundary condition reads
\hbox{$(\sigma\,R\,\partial_R-
\partial_\theta)S\big|_{\theta=\pi/2}=0$}.  Using this boundary
condition and the probability $P$ given by \eqref{P-sol} we deduce
$\sigma\tan\big(\beta\pi\big)=1$ and thus obtain our main result
\eqref{beta-gen}.

\begin{figure}[t]
\includegraphics[width=0.45\textwidth]{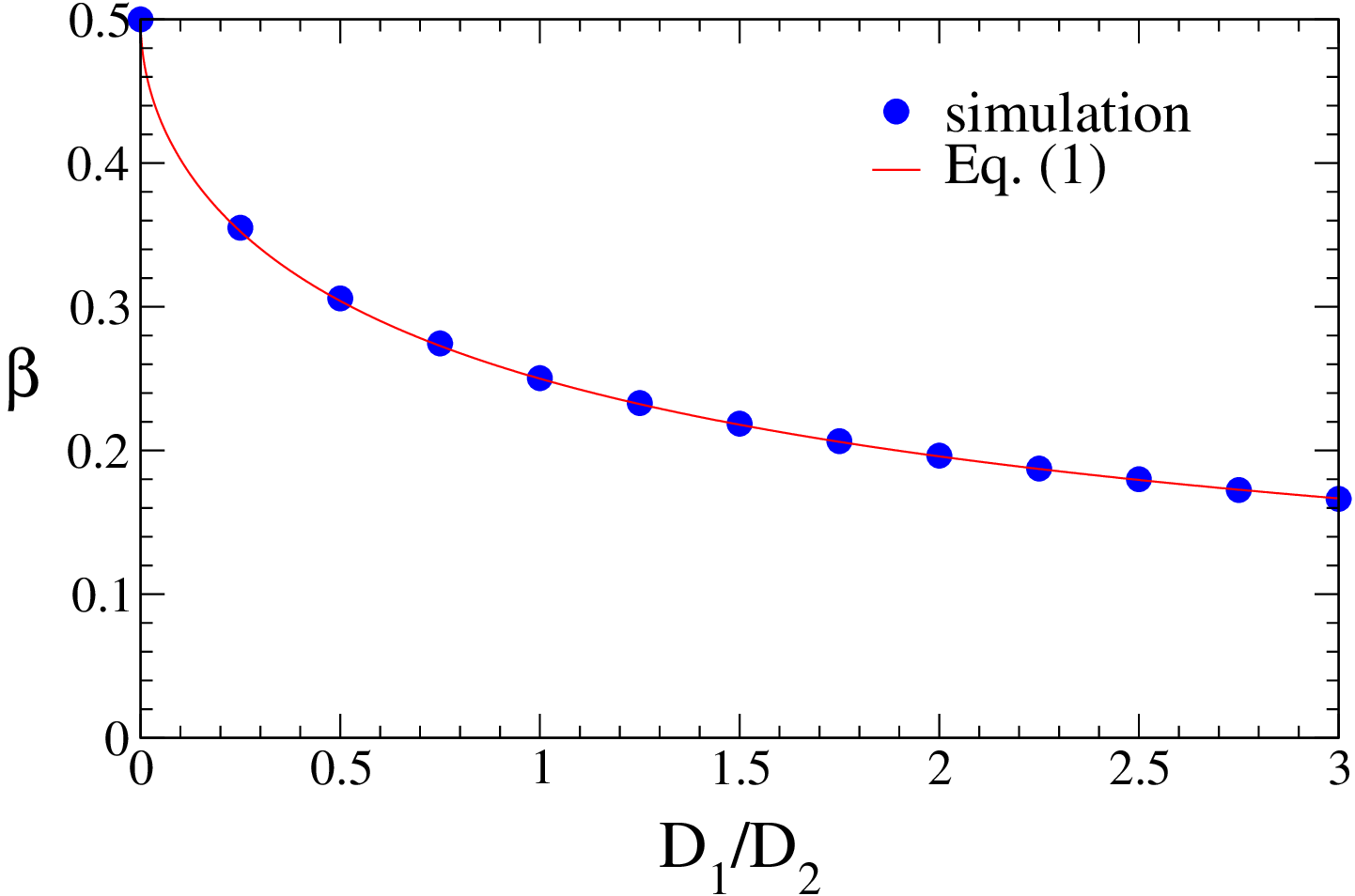}
\caption{The exponent $\beta$ versus the ratio $D_1/D_2$. The line
  corresponds to the theoretical curve \eqref{beta-gen}, and the dots
  to results of Monte Carlo simulations with $10^7$ independent
  realizations.}
\label{fig-beta}
\end{figure}

The exponent is rational for special values of the diffusion
constants, for instance, $\beta(1/3)=1/3$, $\beta(1)=1/4$, and
$\beta(3)=1/6$.  Exponent $\beta$ varies continuously with the ratio
$D_1/D_2$ (Fig.~\ref{fig-beta}). Unlike the universal first-passage
behavior $t^{-1/2}$ characterizing positions of Brownian particles,
the behavior of the probability $P$ is not universal and further, it
can not be derived using heuristic scaling arguments. Further,
first-passage kinetics of maxima of mobile Brownian particles are
generally slower compared with first-passage kinetics of positions
since $\beta<1/2$.  As expected, the limiting values are
$\beta(0)=1/2$ and $\beta(\infty)=0$, and further, the limiting
behaviors are $1/2-\beta \simeq \tfrac{1}{\pi}\sqrt{D_1/D_2}$ when
$D_1\ll D_2$ and $\beta\simeq \tfrac{1}{\pi}\, \sqrt{D_2/D_1}$ for
$D_2\ll D_1$.

One anticipates that the asymptotic behavior \eqref{beta-gen} applies
to a broad class of diffusion processes.  As a test, we performed
Monte Carlo simulations (see also refs.~\cite{pg,obdkgs}) of
discrete-time random walks in one dimension with two different
implementations: (i) a random walk on a lattice where all step lengths
have the same size, and (ii) a random walk on a line where the step
lengths are chosen from a uniform distribution with compact
support. In both cases, the simulation results are in excellent
agreement with the theoretical predictions.  The simulation results
shown in Figures \ref{fig-st} and \ref{fig-beta} correspond to random
walks on a line.

For $n$ particles undergoing Brownian motion, there are three natural
generalizations of the probability $P$.  First is the probability $A_n$
that all $n$ maxima remain perfectly ordered, that is, $n$ staircases
as in Figure \ref{fig-xt} never intersect; for positions of Brownian
particles, this problem dates back to \cite{mef}.  Second is the
probability $B_n$ that rightmost staircase is never overtaken, the
corresponding problem for positions was studied in \cite{bg}.  Third
is the probability $C_n$ that the leftmost staircase never overtakes
another maxima \cite{bjmkr}.  We expect all three quantities to decay
as power laws,
\begin{equation}
\label{abc}
A_n\sim t^{-\alpha_n},\quad B_n\sim t^{-\beta_n},\quad C_n\sim t^{-\gamma_n},
\end{equation}
with exponents $\alpha_n$, $\beta_n$, and $\gamma_n$ that depend on
the number of particles $n$. Table I lists results of Monte Carlo
simulations along with the analogous exponents for the positions,
rather than the maxima \cite{bk-mult}.

\begin{table}[h]
\begin{tabular}{|c|l|l|l|l|l|l|}
\hline 
    & \multicolumn{3}{c|}{maxima} & \multicolumn{3}{c|}{positions}\\
\hline
$n$ & \multicolumn{1}{c|}{$\alpha_n$} & \multicolumn{1}{c|}{$\beta_n$} & \multicolumn{1}{c|}{$\gamma_n$} & \multicolumn{1}{c|}{$a_n$} & 
\multicolumn{1}{c|}{$b_n$} & \multicolumn{1}{c|}{$c_n$}\\
\hline
$3$&$0.653$&$0.432$&$0.335$&\multicolumn{1}{c|}{$3/2$}& \multicolumn{1}{c|}{$3/4$}   & \multicolumn{1}{c|}{$3/8$}   \\
$4$&$1.13$ &$0.570$&$0.376$&\multicolumn{1}{c|}{$3$}  & $0.91$ & $0.306$ \\
$5$&$1.60$ &$0.674$&$0.401$&\multicolumn{1}{c|}{$5$}  & $1.02$ & $0.265$ \\
$6$&$2.01$ &$0.759$&$0.417$&\multicolumn{1}{c|}{$15/2$}  & $1.11$ & $0.234$ \\
\hline
\end{tabular}
\caption{The exponents $\alpha_n$, $\beta_n$, and $\gamma_n$ defined
  in equation \eqref{abc} versus the number of particles $n$.  Also
  shown as a reference are values for the corresponding exponents
  $a_n$, $b_n$, and $c_n$ characterizing analogous probabilities
  involving the positions of $n$ Brownian particles
  \cite{bk-mult}. The only known \cite{mef} sequence of exponents is
  $a_n=n(n-1)/4$.}
\end{table}

All of the exponents are directly related to eigenvalues of the
Laplace operator in high-dimensional space with suitable boundary
conditions.  Even for the simpler case of ordered positions, such
eigenvalues are generally unknown (Table I).  We expect that
\hbox{$\alpha_n>\beta_n>\gamma_n$} and furthermore that all three
exponents increase with $n$. Further, it is possible to justify the
behavior \hbox{$\beta_n \simeq b_n$} and consequently \cite{bk-mult,bjmkr}
obtain the logarithmic growth \hbox{$\beta_n\simeq \tfrac{1}{4}\ln n$}
when the number of particles is large, \hbox{$n\to \infty$}. Also, it
is simple to show that $\gamma_n\to 1/2$ in the limit \hbox{$n\to
\infty$} \cite{bk14}. Based on the numerical results we conjecture
that one of the exponents is rational, $\gamma_n=\tfrac{n-1}{2n}$;
this form is consistent with $\gamma_1=0$ and $\gamma_2=1/4$.

Our results thus far concern diffusion in one spatial dimension, yet
closely related questions can be asked of Brownian motion in arbitrary
dimension $d$. Consider, for example, the maximum distance traveled by
a Brownian particle. If the particle starts at the origin, this
distance equals the radial coordinate in a spherical coordinate
system. We expect that the probability $U_d$ that the maximal radial
coordinate of one particle always exceeds that of another particle
decays algebraically with time, $U_d\sim t^{-\nu_d}$. Our numerical
simulations show that exponent $\nu$ grows rather slowly with
dimension $d$
\begin{equation}
\nu_1=0.563, \qquad \nu_2=0.602, \qquad \nu_3=0.630.
\end{equation}
It would also be interesting to study planar Brownian excursions and
in particular the probability that the convex hull generated by one
particle always contains that of a second particle \cite{bd,mcr}.

We also mention that the first-passage process studied in this letter
is equivalent to a ``competition'' between two records \cite{abn}. As
a data analysis tool, the first-passage probability $P$ is a
straightforward measure and can be used in finance \cite{bp}, climate
\cite{nmt,whk}, and earthquakes \cite{sdt,bk13}.  The notion of
competing maxima could also describe the span of colloidal particles
undergoing simple or anomalous diffusion \cite{csg,lkb}.

In summary, we studied maxima of Brownian particles in one dimension
and found that the probability that such maxima remain ordered decays
as a power law with time. The exponent characterizing this decay
varies continuously with the diffusion coefficients governing the
motion of the particles. When there are two particles, the problem
reduces to diffusion in two dimensions with mixed boundary conditions.
Recent studies show that the eigenvalues characterizing ordering of a
very large number Brownian trajectories obey scaling laws in the
thermodynamic limit \cite{bk-mult} and an interesting open challenge
would be to use such scaling methods to elucidate extreme value
statistics of many Brownian trajectories.

\smallskip
%\acknowledgements 

We acknowledge DOE grant DE-AC52-06NA25396 for support (EB).

\newpage
\appendix
\onecolumngrid
\section{Supplemental Material}

To derive the various boundary conditions, we consider two discrete
time random walks.  We implement the random walk process as follows:  
in each time step one of the random walks is chosen at random and it
jumps left or right with equal probabilities. Therefore, 
the coordinates $x$ and $y$ evolve according to 
\begin{equation}
\label{rw-xy}
(x,y)\to 
\begin{cases}
(x+1,y)& {\rm with\ prob.\ }1/4;\\
(x-1,y)& {\rm with\ prob.\ }1/4;\\
(x,y+1)& {\rm with\ prob.\ }1/4;\\
(x,y-1)& {\rm with\ prob.\ }1/4.
\end{cases}
\end{equation}
Consequently, the distances $u$ and $v$ change as follows 
\begin{equation}
\label{rw-uv}
(u,v)\to 
\begin{cases}
(u+1,v)& {\rm with\ prob.\ }1/4,\\
(u-1,v)& {\rm with\ prob.\ }1/4,\\
(u,v+1)& {\rm with\ prob.\ }1/4,\\
(u,v-1)& {\rm with\ prob.\ }1/4,
\end{cases}
\end{equation}
when $u>0$ and $v>0$.  To derive the diffusion equation $\partial_t  
\rho =D\nabla^2\rho$ with $D=1/4$ we write the
recursion equation
\begin{equation}
\label{rho-rec}
\rho(u,v,t+1)=\frac{\rho(u+1,v,t)+\rho(u-1,v,t)+\rho(u,v+1)+P(u,v-1,t)}{4}.
\end{equation}
We now expand the left-hand as a first-order Taylor expansion in time
and the right-hand side as a second-order Taylor series in space.  The
diffusion equation is subject to the boundary conditions
\begin{equation}
\label{bc-uv}
\left(\frac{\partial \rho}{\partial u}-\frac{\partial \rho}{\partial
v}\right)\Big|_{u=0}=0 \qquad\text{and}\qquad \rho\big|_{v=0}=0.
\end{equation}
These boundary conditions follow from the jump rules along the lines
$u=0$ and $v=0$ respectively,
\begin{equation}
\label{rw-boundaries}
(0,v)\to 
\begin{cases}
(0,v+1)& {\rm with\ prob.\ }1/2;\\
(0,v-1)& {\rm with\ prob.\ }1/4;\\
(1,v)  & {\rm with\ prob.\ }1/4.
\end{cases}
\qquad
\text{and} 
\qquad
(u,0)\to 
\begin{cases}
(u+1,0)& {\rm with\ prob.\ }1/4;\\
(u-1,0)& {\rm with\ prob.\ }1/4;\\
(u,1)  & {\rm with\ prob.\ }1/4.
\end{cases}
\end{equation}
In the second case, with probability $1/4$, the second random random walk tries
to overtake the maximum set by the first walker and thus, it is
annihilated (absorbed by the boundary $v=0$).

As function of the initial coordinates $X$ and $Y$, the probability
$P(X,Y,t)$ satisfies a recursion equation analogous to \eqref{rho-rec}
\begin{equation}
\label{P-rec}
P(X,Y,t+1)=\frac{P(X+1,Y,t)+P(X-1,Y,t)+P(X,Y+1,t)+P(X,Y-1,t)}{4}, 
\end{equation}
leading to the diffusion equation for $P$ with $D=1/4$.  The boundary
condition \hbox{$(\partial P/\partial X+\partial P/\partial
Y)\big|_{X=0}=0$} follows from the recursion 
\begin{equation}
P(0,Y,t+1)=\frac{P(1,Y,t)+P(0,Y-1,t)+2P(0,Y+1,t)}{4}.
\end{equation}
To derive this equation, we have to take into account all relevant
initial conditions \hbox{$(m_0,x_0,y_0)$} where $m_0$ is the initial
maximum.  The first term corresponds to \hbox{$(x_0+1,x_0,y_0)$}, the
second to \hbox{$(x_0,x_0,y_0+1)$}, and the third includes
contributions from two initial conditions: \hbox{$(x_0,x_0,y_0-1)$}
and \hbox{$(x_0+1,x_0+1,y_0)$}. Of course, we are interested in the
behavior along the line $X=0$ which corresponds to $m_0=x_0$, but the
problem is well defined for all $m_0\geq x_0\geq y_0$, a region which
lies entirely inside the first quadrant ($X\geq 0$ and $Y \geq 0$) in
the $X$-$Y$ plane. Finally, we stress that in our reduced two-variable
description, we ``integrate'' over the initial maximum $m_0$.


\begin{thebibliography}{99}

\bibitem{mp} P. M\"orders and Y. Peres, {\it Brownian Motion}
      (Cambridge University Press, Cambridge, 2010).

\bibitem{sa}
       E.~Sparre Andersen, 
       Math. Scand. {\bf 1}, 263 (1953); {\it ibid}. {\bf 2}, 195 (1954). 

\bibitem{wf} 
      W.~Feller, 
      {\it An Introduction to Probability Theory and Its Applications}
      (Wiley, New York, 1968).

\bibitem{sr} 
      S.~Redner, {\it A Guide to First-Passage Processes} (Cambridge
      University Press, Cambridge, 2001).

\bibitem{pl}
      P.~L\'evy, 
      {\it Processus Stochastiques et Mouvement Brownien} 
      (Gauthier-Villars, Paris, 1948).

\bibitem{im} 
      K.~It\^{o} and H.~P.~McKean, 
      {\it Diffusion Processes and Their Sample Paths} 
      (New York, Springer, 1965).
      
\bibitem{msb}
        S.~N.~Majumdar, C.~Sire, A.~J.~Bray, and S.~J.~Cornell, 
	Phys. Rev. Lett. {\bf 77}, 2867 (1996).

\bibitem{dhz} 
        B.~Derrida, V.~Hakim, R.~Zeitak, 
	Phys. Rev. Lett. {\bf 77}, 2871 (1996).

\bibitem{to} 
       Y.~Takikawa and H.~Orihara, 
       Phys. Rev. E {\bf 88}, 062111 (2013).

\bibitem{csg}
      G.~Coupier, M.~Saint Jean, C. Guthmann, 
      EPL {\bf 77}, 60001 (2007). 

\bibitem{bk}
      E.~Ben-Naim and P.~L.~Krapivsky, 
      Phys. Rev. Lett. {\bf 102}, 190602 (2009). 

\bibitem{lkb}
     C.~Lutz, M.~Kollmann, and C.~Bechinger,  
     Phys. Rev. Lett. {\bf 93}, 026001 (2004).

\bibitem{bs}
      E.~Barkai and R.~Silbey, 
      Phys. Rev. Lett. {\bf 102}, 050602 (2009). 

\bibitem{krb}
      P. L. Krapivsky, S. Redner, and E. Ben-Naim,
      {\it  A Kinetic View of Statistical Physics}
      (Cambridge University Press, Cambridge, 2010).

\bibitem{bms}
      A.~J.~Bray, S.~N.~Majumdar, and G.~Schehr, 
      Adv. Phys. {\bf 62}, 225 (2013).      

\bibitem{dls}
      B.~Derrida, J.~L.~Lebowitz, and E.~R.~Speer, 
      Phys. Rev. Lett. {\bf 15}, 150601 (2001). 

\bibitem{smcrf}
      G.~Schehr,  S.~N.~Majumdar, A. Comtet, and J.~Randon-Furling,       
      Phys. Rev. Lett. {\bf 101}, 150601 (2008). 

\bibitem{mz}
      S.~N.~Majumdar and R.~M.~Ziff,
      Phys. Rev. Lett. {\bf 101}, 050601 (2008).

\bibitem{bk14}  
      E.~Ben-Naim and P.~L.~Krapivsky, 
      arXiv:1404.2966.

\bibitem{ghw} 
      G.~H.~Weiss, 
      {\it Aspects and Applications of the Random Walk} 
      (North-Holland, Amsterdam, 1994).

\bibitem{bk-mult}
      E.~Ben-Naim and P.~L.~Krapivsky,  
      J. Phys. A {\bf 43}, 495007 (2010); J. Phys. A {\bf 43}, 495008 (2010).

\bibitem{cr}
      D.~Considine and S.~Redner,
      J. Phys. A {\bf 22}, 1621 (1988).

\bibitem{pg}
      P.~Grassberger,
      Computer Phys. Comm. {\bf 147}, 64 (2002).

\bibitem{obdkgs}
      T.~Oppelstrup, V.~V.~Bulatov, A.~Donev, M.~H.~Kalos,
      G.~H.~Gilmer, and B.~Sadigh,
      Phys. Rev. E {\bf 80}, 066701 (2009).

\bibitem{mef}
      M.~E.~Fisher, J.~Stat. Phys. {\bf 34}, 667 (1984);
      D.~A.~Huse and M.~E.~Fisher,
      Phys. Rev. B {\bf 29}, 239 (1984).

\bibitem{bg} M. Bramson and D. Griffeath,
     in: {\it Random Walks, Brownian Motion, and Interacting
     Particle Systems: A Festshrift in Honor of Frank Spitzer}, eds.\
     R. Durrett and H. Kesten (Birkh\"auser, Boston, 1991).

\bibitem{bjmkr}
      P.~L.~Krapivsky and S.~Redner, J.\ Phys.\ A {\bf 29}, 5347 (1996); 
      D.~ben-Avraham, B.~M.~Johnson, C.~A.~Monaco, P.~L.~Krapivsky,
      and S.~Redner,  J.~Phys.~A {\bf 36}, 1789 (2003). 

\bibitem{bd} 
        B.~Duplantier, in 
 	{\it Fractal Geometry and Applications: A Jubilee of Beno\^{i}t Mandelbrot} 
        (M. L. Lapidus and M. van Frankenhuysen, eds.), 
        Proc. Symposia Pure Math. {\bf 72}, 365 (2004). 

\bibitem{mcr}
      S.~N.~Majumdar, A.~Comtet, and J.~Randon-Furling,              
      J. Stat. Phys. {\bf 138}, 955 (2010).

\bibitem{abn}
       B.~C.~Arnold, N.~Balakrishnan and H.~N.~Nagraja, 
       {\it Records} (Wiley-Interscience, 1998). 

\bibitem{bp}
      J.~-P.~Bouchaud and M.~Potters, 
      {\em Theory of Financial Risk and Derivative Pricing}
      (Cambridge University Press, Cambridge 2003).

\bibitem{nmt}
      W.~I.~Newman, B.~D.~Malamud, and D.~L.~Turcotte, 
      Phys. Rev. E {\bf 82}, 066111 (2010).

\bibitem{whk} 
      G.~Wergen, A.~Hense, and J.~Krug, 
      Climate Dynamics {\bf 42}, 1275 (2014).

\bibitem{sdt} 
       R.~Shcherbakov, J.~Davidsen, and K.~F.~Tiampo, 
       Phys. Rev. E {\bf 87}, 052811 (2013).

\bibitem{bk13}  
      E.~Ben-Naim and P.~L.~Krapivsky, 
      Phys.\ Rev.\ E {\bf 88}, 022145 (2013); 
      P.~W.~Miller and E.~Ben-Naim, 
      J. Stat. Mech. P10025 (2013).

\end{thebibliography}
\end{document}